# Classification of anatomic structures in head and neck by CT-based radiomics

Yoichi Watanabe[1], A. Biswas[2], K. Rangarajan[2], G. Rath[2], and N. Gopishankar[2].

1 University of Minnesota, Minneapolis, MN, USA

2 All India Institute of Medical Sciences, New Delhi, India

**Abstract**

**Background and Purpose:** Radiomics features are used to identify disease types and predict therapy outcomes. However, how the radiomics features are different among different anatomical structures has never been investigated. Hence, we analyzed the radiomics features of 22 anatomical structures in the head and neck area in CT images. Furthermore, we studied whether CT radiomics can classify anatomical structures of the head and neck using unsupervised machine-learning techniques.

**Materials and methods:** We obtained IMRT/VMAT treatment planning data from 36 patients treated for head and neck cancers in a single institution. There were 1357 contours of more than 22 anatomical structures drawn on planning CTs. We calculated 174 radiomics features using the SIBEX program. First, we tested whether the radiomics features of anatomical structures were unique enough to classify all contours into 22 groups. We then developed a two-stage clustering technique to classify 22 anatomic structures into sub-groups with similar physiological or biological characteristics.

**Results:** The heatmap of 174 radiomics features of 22 anatomical structures showed a distinct difference among tumors and other healthy structures. Radiomics features have allowed us to identify the eyes, lens, submandibular, pituitary glands, and thyroids with over 90% accuracy. The two-stage clustering of 22 structures resulted in six subgroups, which shared common characteristics such as fatty and bony tissues.

**Conclusions:** We have shown that anatomical structures in head and neck tumors have distinguishable radiomics features. We could observe similarities of features among subgroups of the structures. The results suggest that CT radiomics can help distinguish the biological characteristics of head and neck lesions.

Abbreviations

| | |
|---|---|
| CT | Computed tomography |
| CTV | Clinical target volume |
| FBN | Fixed bin number |
| FBS | Fixed bin size |
| GTV | Gross tumor volume |
| PCA | Principal component analysis |
| PC | Principal component |
| PTV | Planning target volume |
| WCSS | Within cluster sum of squares |
| TSS | Total cluster sum of squares |
| BCSS | Between cluster sum of squares |
| $\Lambda$ | The ratio of BCSS/TSS |

Corresponding Authors

Gopishankar Natanasabapathi, PhD,

Assistant Professor of Radiotherapy (Medical Physics),

Department of Radiation Oncology, Dr. B. R. A. IRCH,

All India Institute of Medical Sciences, Ansari Nagar,

New Delhi - 110029, India. Cell No - +919650553660

gshankar1974@yahoo.com

**Introduction**

Radiomics is increasingly used for quantitative characterization of diagnostic images, particularly as potential biomarkers for evaluating cancer therapy, including radiation therapy. Many applications of radiomics are focused on predicting the cancer therapy outcome and toxicity with the radiomics features as biomarkers or predictors (1-14). Furthermore, there are many studies in which radiomics features were used to differentiate tumor from healthy tissue (15, 16), classify the tumor type (17, 18), and determine the tumor grade (19-21). Some investigated differences in radiomics features of tumors in different organs (22). Applications of radiomics to discover the genomics from diagnostic images are another active area, and it is called radiogenomics (23). In addition, there are emerging applications for detecting microenvironments such as hypoxia (24, 25) and studying tumor immune biology (26).

Despite the immense activities in radiomics applications in radiology and radiation oncology, as evidenced by the number of publications on the topic, to the author's knowledge, no one investigated the differences of radiomics features among different organs and structures, including tumor lesions. Hence, we studied the variability of radiomics feature values by analyzing the radiomics features of anatomical structures in head and neck areas. In this study, we addressed three questions:

1. Are radiomics features of tumors different from those of healthy structures in the head and neck area?
2. Do anatomical structures and tumors have structure-specific radiomics features?
3. Can anatomical structures be classified into subcategories based on their functions and tissue types using radiomics?

To answer question #3, we needed to categorize M anatomical structures in the head and neck area into J categories or structure groups, which may share common physiological and biological characteristics using radiomics features of N samples obtained from treatment planning CT image data. In this study, we solved these problems by an unsupervised machine learning (ML) technique, i.e., clustering algorithms. A simple application of the algorithm could not classify all contours into J categories. Hence, we developed a two-stage clustering method to accomplish the goal. In Stage-1, we obtained six clusters or subgroups from N samples or contours belonging to one of M anatomical structures. Using the known anatomical structure of those contours, we calculated the probabilities that a contour is assigned to one of six subgroups. Using the probabilities as parameters for the stage-2 clustering, we classified M anatomical structures into J categories.

The results will demonstrate the ability of radiomics in differentiating the biological processes taking place in different parts of the head and neck having various types of tissue, cellular structures, and physiological functions, which are hidden under the medical images or CT images in this study. Hence, this study will help us enhance our confidence in the radiomics features as potential biomarkers of treatment response and radiation toxicity.

**Materials and Methods**

*Patient data*

We obtained the CT image data sets for 36 patients treated in a single institution, the All India Institute of Medical Sciences (AIIMS) in Delhi, India, for head and neck cancers using the IMRT/VMAT technique. The demographic data and tumor locations of these patients are summarized in Table 1. All patients were scanned using Philips Brilliance Big Bore CTs with the following scanning parameters: helical, 120kVp, 450mAs, 512x512 pixels, slice thickness: 1, 2, or 3mm, FOV: between 240 mm x240 mm and 600 mm x 600 mm, and 16 bits per pixel. For the current study, we chose 22 anatomical structures, including 92 gross tumor volumes (GTV), a total of 67 right and left parotids, and 35 mandibles. See Table 2 for the complete list. There were 1357 contours, but for the current study, we chose 1175 contours of 22 anatomical structures mentioned above. Experienced radiation oncologists segmented all structures for treatment planning. See Figure 1 for an example of CT images showing the contours of GTV, right and left parotids, and mandibles.

Table 1: Patient characteristics (36 patients)

| Age | | 50.9 ±16.79 [18-86] |
|---|---|---|
| Sex | Female | 13 |
| | Male | 23 |
| Primary region | Nasopharynx | 10 |
| | Nasal cavity/Maxillary sinus | 8 |
| | Neck | 6 |
| | Tongue/Base of tongue | 5 |
| | Salivary glands (parotid/submandibular) | 3 |
| | Tonsil | 2 |
| | Others (cheek/skull) | 2 |

Table 2: Characteristics of structures

| No. | Structure Name | Function | Tissue/cell type | $N_i$ [1] | Volume [voxels] | CT number |
|---|---|---|---|---|---|---|
| 1 | GTV | Tumor | Cancer | 92 | 52454 | 25.9 |
| 2 | CTV | Tumor | Cancer, others | 84 | 195608 | 27.9 |
| 3 | PTV | Tumor | Cancer, others | 76 | 389018 | 39.3 |
| 4 | Parotids | Salivary gland | Serous, ductal | 67 | 24476 | 5.8 |
| 5 | Mandibles | Support structure | Bone | 35 | 57672 | 36.5 |
| 6 | Eyes | Vision | Water, connective | 66 | 8292 | 4.5 |
| 7 | Lens | Vision | Lens fibers, epithelial | 65 | 205 | 2.1 |
| 8 | Lacrimal glands | Secretory | Serous, ductal | 63 | 695 | 4.8 |
| 9 | Optic nerves | Vision | Nerve fibers | 72 | 903 | 4.3 |
| 10 | Cochleae | Auditory | Bone | 66 | 188 | 27.8 |
| 11 | Spinal cord | CNS | Nervous | 57 | 32257 | 7.0 |
| 12 | Chiasm | Vision | Nerve fibers | 32 | 943 | 1.9 |
| 13 | Brainstem | CNS | Neural | 55 | 31727 | 2.4 |
| 14 | Temporal lobe | CNS | Neural | 64 | 102592 | 2.9 |
| 15 | Submandibular | Salivary gland | Serous, mucous | 53 | 7948 | 5.3 |
| 16 | Lips | Respiratory | Squamous epithelium | 30 | 22647 | 18.4 |
| 17 | Esophagus | Digestion | Squamous epithelium | 30 | 9185 | 26.5 |
| 18 | Pituitary gland | Endocrine | Chromophils, neurosecretory | 28 | 347 | 2.6 |

| 19 | Thyroid gland | Endocrine | Follicular | 27 | 9509 | 7.2 |
| 20 | Larynx | Respiratory | Cartilage, columnar epithelium | 25 | 28442 | 30.6 |
| 21 | Oral cavity | Digestion | Squamous epithelium | 32 | 73133 | 23.8 |
| 22 | TMJ | Structure | Bone | 56 | 2527 | 18.1 |

(1) The total number of contours belonging to the i-th anatomical structure.

Figure 1: CT images of a typical patient with the contours of GTV, right and left parotids, and mandible.

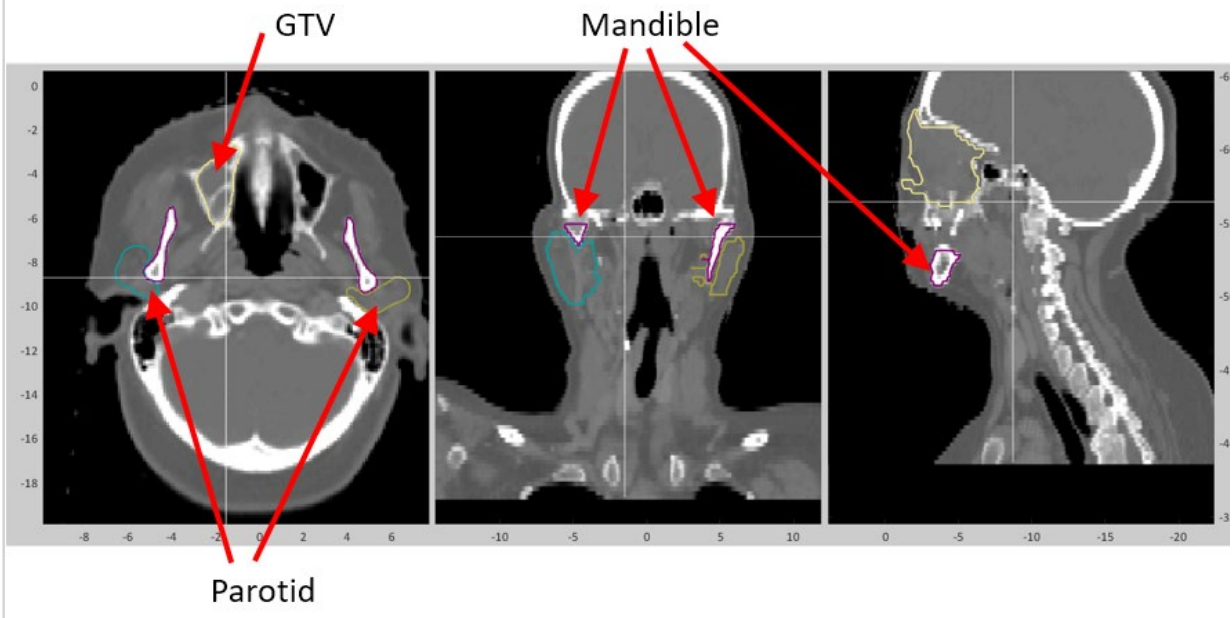

*Radiomics*

We calculated 174 radiomics feature values by the SIBEX software (27), which complies with the recommendations of the International Biomarker Standardization Initiative (IBSI) (28). There are 11 categories of radiomics features, as shown in Table 3, which lists the number of features in each category. The complete list of all features can be found in Appendix A. The details of feature calculation algorithms are discussed in the IBSI report (28). The radiomics calculations were done using three processing parameters: Fixed Bin Number (FBN) with 32 bins (FBN32) and Fixed Bin Size (FBS) with 25 bin widths (FBS25) and ten bin widths (FBS10).

Table 3: Category and the number of features used for the radiomics analysis. The total number of features was 174.

| Category number | Category name | The number of features |
|---|---|---|
| 1 | GrayLevelCoocurrenceMatrix | 25 |
| 2 | GrayLevelDistZoneMatrix | 16 |
| 3 | GrayLevelRunLengthMatrix | 16 |
| 4 | GrayLevelSizeZoneMatrix | 17 |
| 5 | IntensityDirect | 17 |
| 6 | IntesnityHistogram | 23 |
| 7 | IntensityVolumeHistogram | 7 |
| 8 | LocalIntensityFeatures | 2 |
| 9 | MorphologicalFeatures | 29 |
| 10 | NeighborGLDependence | 17 |
| 11 | NeighborIntensityDifference | 5 |

*Clustering*

The feasibility of clustering 1175 contours of 22 anatomical structures into subgroups was studied in four steps, as summarized in the process diagram of Figure 2. In the first step (the second from the top), the number of radiomics features used for clustering was reduced from 174 radiomics features to *p* composite features using the principal component analysis (PCA) method (29). In the second step (the stage-1 clustering), we used a clustering algorithm to assign 1175 contours to one of the K clusters or subgroups. Here the contours of an anatomical structure could be assigned to more than one subgroup. By counting the number of contours, $n_{i,k}$ of a known structure, *i*, which was assigned to a subgroup, *k*, by the clustering, we can generate a frequency distribution of the number of contours of a structure assigned to the subgroups. In step 3, we calculated the normalized frequency, or probability, of contours of anatomical structure, i, assigned to a subgroup k, $P_{i,k}$, by

$$P_{i,k} = \frac{n_{i,k}}{N_i} \quad (1)$$

Here, $N_i$ is the number of contours known to belong to the structure *i*. So, for every anatomical structure *i*, the following equation holds:

$$\sum_{k=1}^{K} P_{i,k} = 1 \quad (2)$$

Considering $P_{i,k}$ as a set of new features for the 22 anatomical structures, we did the stage-2 clustering in step 4. The results were 22 anatomical structures categorized into L structure groups, representing tissue type or other similarities among the structures in the same structure group.

Figure 2: Process map of the analysis procedure

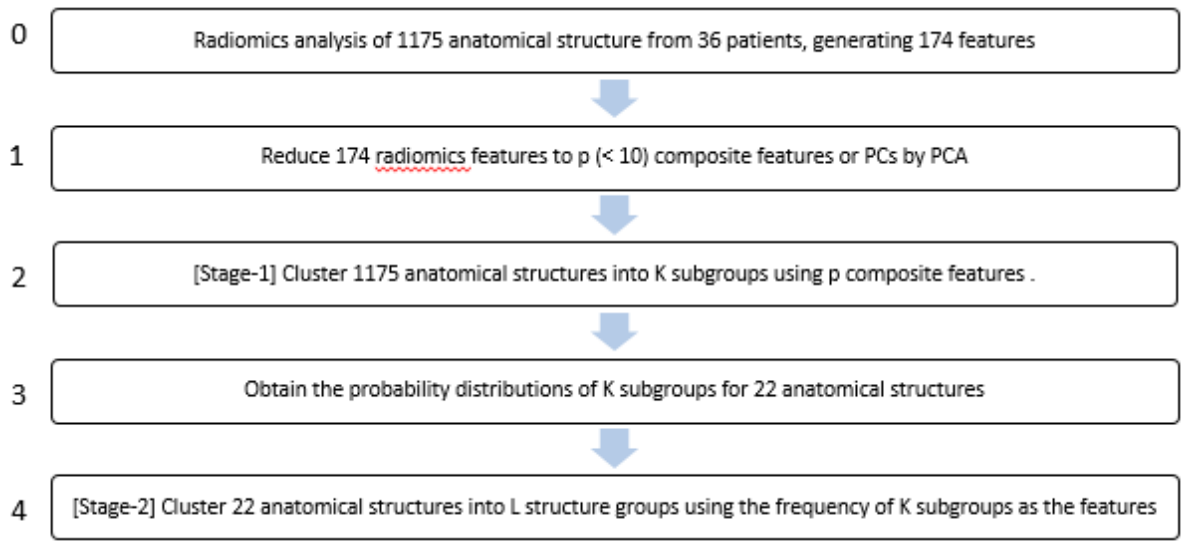

For the clustering analysis, radiomics feature values were normalized using the Z-score or linear range methods (29). For the latter, the values were scaled to the range of 0 and 1. We tested varying numbers of principal components p, i.e., p = 2 to 174, for the stage-1 clustering analysis.

After step 1, we generated a heatmap, which presented the distributions of 174 radiomics features for all 1175 contours. Then, the stage-1 clustering was done using the k-means algorithm (29). Next, we evaluated the clustering performance by varying the number of subgroups K from 2 to 22. To assess the performance of the stage-1 clustering for K=22, we calculated the clustering accuracy of 22 anatomical structures. The stage-2 clustering was made using two clustering methods: the k-means and the hierarchical clustering, which generated dendrograms (29). Additionally, we tested the consensus clustering method (30, 31) as an alternative to the two-stage clustering method. We did several tests to select optimal hyper-parameters and algorithms by changing analysis options and parameter values for the sensitivity of the final results on the variations. Note that there is the following relationship between the within-cluster sum of squares of cluster k ($WCSS_k$), the between-cluster sum squares (BCSS), and the total cluster sum of squares (TSS):

$$TSS = BCSS + \sum_{k=1}^{K} WCSS_k \quad (3)$$

Note that the second term on the right side of Eq. (3) is the total between-cluster sum of squares (TWCCS). We defined the ratio of the between-cluster sum squares (BCSS) and the total cluster sum of squares (TSS), denoted by $\Lambda$:

$$\Lambda = \frac{BCSS}{TSS} \quad (4)$$

The $\Lambda$ value in percentage varied from 0 to 100%. Note that a smaller $WCSS_k$ means tighter clustering of its members in cluster k. Hence, a larger $\Lambda$ implies a better clustering. The k-means routine of the R package prints $WCSS_k$ and $\Lambda$ (32). Therefore, the performance with different calculation parameters could be evaluated by TWCCS or $\Lambda$.

**Results**

*Selection of parameters and methods*

Table 4 shows the parameter values and methods used to optimize the analysis method for the data presented in this study. The binning method used to calculate the radiomics features affects the calculated absolute values. We tested the fixed bin size (FBS) with 10 bins and 25 bins and the fixed bin number (FBN) with 25 bins. Based on our test results and a recommendation made in Bettineli et al.(27), we chose the FBS with 25 bins (FBS25).

Before the principal component analysis, we scaled the radiomics feature values by the Z-score and Linear-range methods. We found that the former resulted in a slightly better clustering accuracy $\Lambda$ than the latter. Hence, we decided to use the Z-score scaling method for the current study.

There should be a sufficient number of principal components (PC) to closely represent the parameter space represented originally by 174 radiomics features. Since the smaller number of PC is desirable for the subsequent clustering analysis, we examined the results by varying the number of PCs, p, from 2 to 174. We found 5 PCs could model 70% of data points and decided to use this number for the rest of the study. The relevant data will be presented in the later section.

The cluster analysis such as the k-means method, requires the user to preselect the number of clusters, K. Ideally, there should be 22 clusters since there are 22 anatomical structures. However, as shown later in Figure 5, the clustering accuracy represented by the parameter TWCSS decreases as K increases and eventually becomes less sensitive to K. Hence, in the following analysis, we set K to 6.

We tested three methods among many clustering techniques: k-mean and hierarchical for the stage-2 clustering. Our tests indicated that these methods gave the same results. Hence, we used the hierarchical clustering method for the rest of the study, which was more robust and informative than the k-means method. A significant disadvantage with the k-means algorithm is its need to preselect the K value.

Table 4: Alternative analysis parameters and methods. The items in bold were selected for the remainder of the analyses.

| Method or parameters | Tested options/methods/values |
|---|---|
| Bin type and numbers for radiomics feature calculations | FBS10, **FBS25**, FBN 32 |
| Scaling of radiomics feature values | **Z-score** <br> Linear-range |
| The number of principal components, $p$ | 2 to 174, **5** |
| The number of subgroups, K | 2 to 22. **6** and **22** |
| Clustering methods for Stage-2 | k-means <br> **hierarchical** <br> consensus clustering |

*Radiomics features of anatomical structures*

Table 2 lists the average volumes and average CT numbers of 22 anatomical structures. These were two of the radiomics features calculated by the SIBEX program (MorphologicalFeatures-Volume and IntensityDirect-Mean). The table also includes the function and tissue type of each anatomical structure.

The heatmap in Figure 3 shows the quantitative distribution of 174 radiomics features among all 1357 contours, which were reordered according to anatomical structures. One can visually confirm that some structures have different radiomics feature values from others. For example, 252 contours from the bottom of the diagram are GTV, CTV, or PTV. The parotids are from 253-319, and the mandibles are from 320-354. Parotid and mandible had radiomics values distributed very differently from GTV, CTV, and PTV. For example, the differences can be more clearly observed in Figures 4 (a), (b), and (c), which show the plots of radiomics feature values of the esophagus and oral cavity in comparison to GTV. A data point on the solid oblique line in Figure 4 means that two structures in the x and y-axes had the same feature values. The distributions of points, each of which was one contour, showed apparent differences, but to different magnitudes among the anatomical structures.

Figure 3: Heatmap. 1357 anatomical contours in the horizontal axis and 174 radiomics features in the vertical axis.

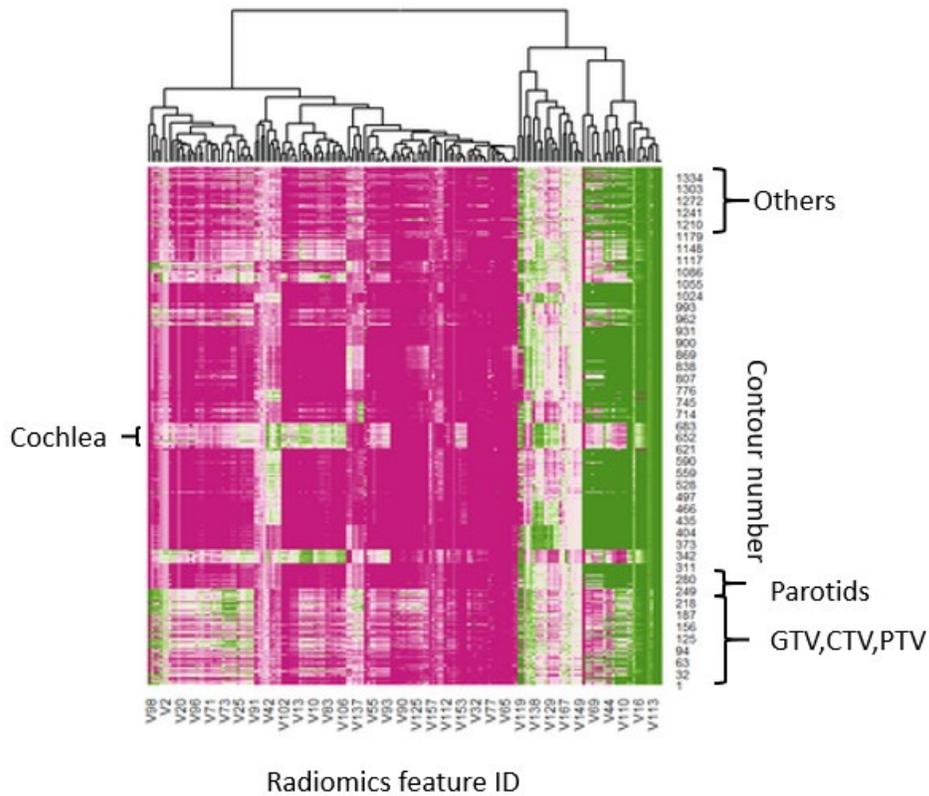

Figure 4: Correlation of radiomics feature values

(a) GTV (X) vs. Esophagus (Y)   (b) GTV (X) vs. Oral cavity (Y)   (c) GTV (X) vs. Larynx (Y)

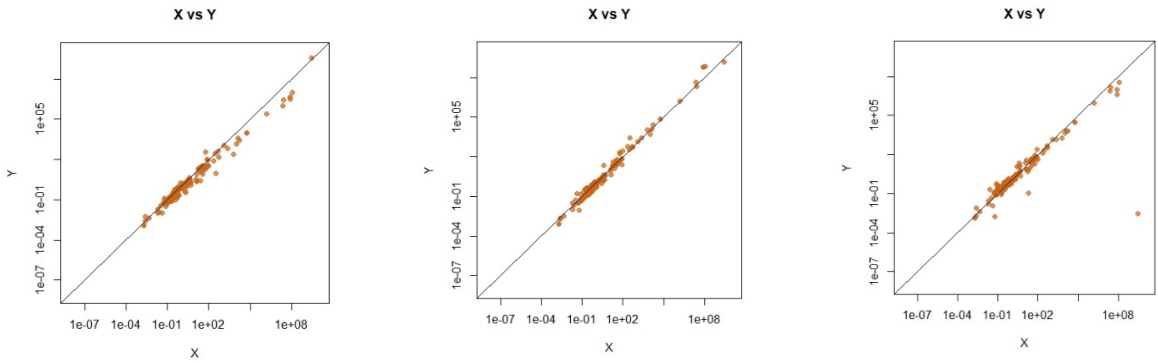

Figure 5: Total within clusters sum of squares (TWCCS) vs. the number of clusters

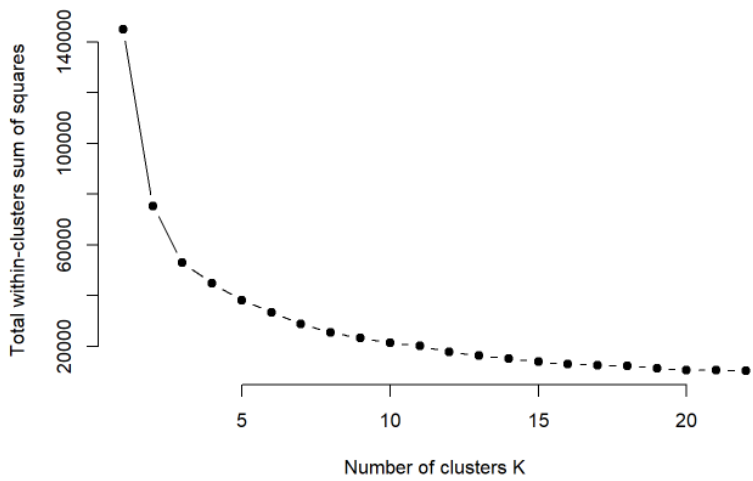

*PCA*

A scatter plot of 1175 contours using the first two principal components, PC1, and PC2, is shown in Figure 6, where different colors indicate different anatomical structures. We can observe weak, but recognizable concentrations of some structures in a specific area in the figure. For example, GTV indicated by black dots are in the middle PC2 range, although those points widely spread out in the PC1 direction. PTV showed a distribution similar to GTV but in the lower PC2 area. Cochleae and mandible were clustered in specific regions of the plot.

Figure 6: 1175 contours plotted for the axes with the first two principal components

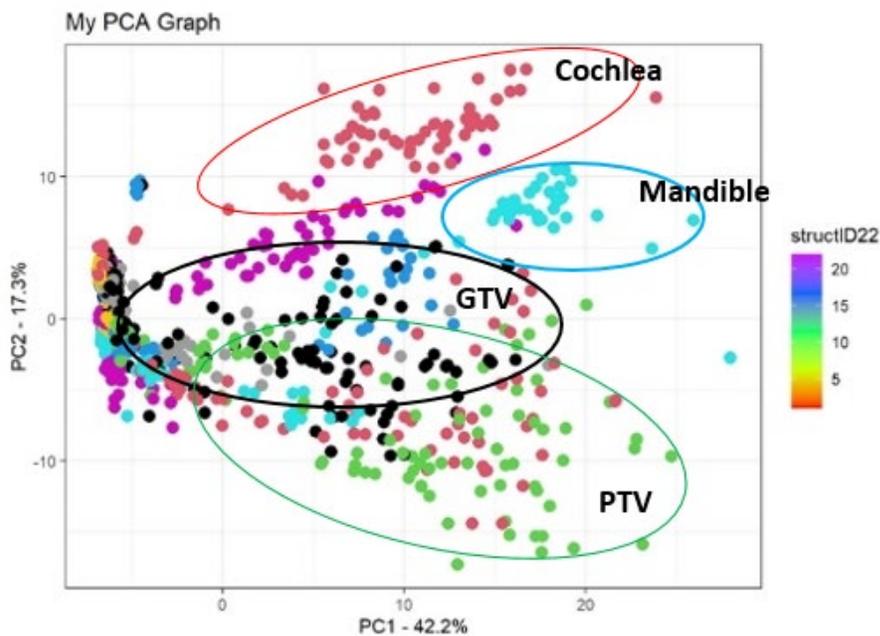

The scree plot in Figure 7 shows the percent variations or the proportion of variance explained (PVE) as a function of the number of the principal components. The graph indicates that the first few components may be sufficient for clustering contours. In particular, the five largest components accounted for 74% of the variation (or the cumulative proportion of 0.74). Hence, we chose five PCs for the stage-1 clustering analyses.

Figure 7: Scree plot: Percent Variation vs. the number of principal components

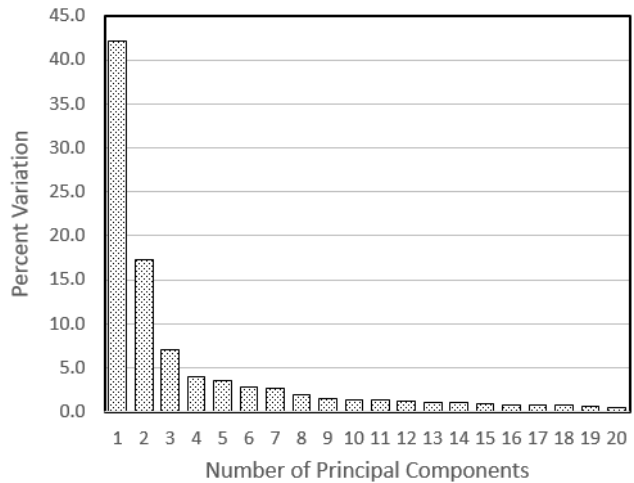

The PCA program ranked the radiomics features based on the importance of the radiomics features for good clustering. Table 5 shows the ten most important features. It is noted that the 7th feature was IntensityDirect/Mean, which was the mean CT numbers.

Table 5: Top 10 radiomics feature influencing PCs

| Rank | Feature ID no. | Feature name |
| --- | --- | --- |
| 1 | 81 | IntensityDirect/ 90th Percentile |
| 2 | 117 | IntensityVolumeHistogram/ IntensityFolFrac_10 |
| 3 | 91 | IntensityDirect/ RootMeanSquare |
| 4 | 55 | GrayLevelRunLengthMatrix/ GrayLevelVariance |
| 5 | 51 | GrayLevelRunLengthMatrix/ GLNonuniformityNorm |
| 6 | 99 | IntensityHistogram/ 90thPercentile |
| 7 | 74 | IntensityDirect/ Mean |
| 8 | 45 | GrayLevelRunLengthMatrix/ HighGLRunEmpha |
| 9 | 47 | GrayLevelRunLengthMatrix/ ShortRunHighGLEmpha |
| 10 | 156 | NeighborGLDependence/ HighGLCountEmpha |

*Clustering of anatomical structures*

<u>*Stage-1*</u>

First, the 1175 contours were clustered into 22 subgroups (or K=22) with the k-means clustering method with five PCs. Ideally, all elements or contours of each anatomical group should be in the same subgroup. Figure 8 presents the clustering results, where the horizontal axis indicates the 22 anatomical structures, and the vertical axis indicates the 22 subgroups or cluster groups. The red color shows almost all contours are assigned to a specific group. For example, the box at the crossing of structure #4 (parotids) and subgroup 4 has a reddish color, implying a sound performance of the algorithm in assigning the parotid contours to this subgroup. In fact, 53 out of 67 parotid contours were assigned to subgroup 4 (79.1% accuracy). Structure #5 (mandible) had an even higher accuracy with 30 out 35 correct assignments, or 85.7% accuracy. Contours with an accuracy higher than 90% were structures #6 (Eyes), #7 (Lens), #15 (Submandibular), #18 (Pituitary gland), and #19 (Thyroid gland). See Table 6 for the complete list. It is noted that the contours of tumors such as GTV, CTV, and PTV, were assigned to multiple subgroups. Some subgroups had contours belonging to several anatomical structures. For example, subgroup 9 included the contours of structure #8 (Lacrimal glands), #9 (Optic nerves), and #12 (Chiasm).

Figure 8: Accuracy of clustering of 1175 contours of 22 anatomical structures to 22 subgroups

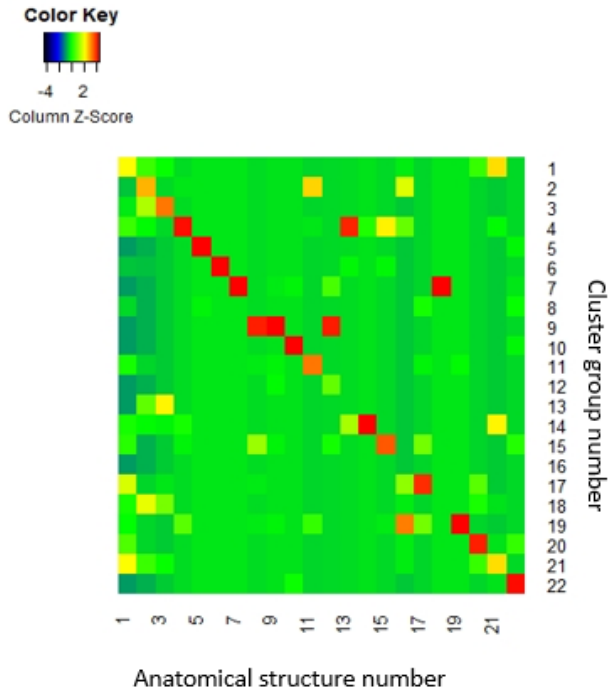

Table 6: Classification accuracy by k-means method with 22 clusters and PC#=5

| No. | Structure Name | Maximum %ratio |
|---|---|---|
| 1 | GTV | 21.7% |
| 2 | CTV | 25.0% |
| 3 | PTV | 26.3% |
| 4 | Parotids | 79.1% |
| 5 | Mandibles | 85.7% |
| 6 | Eyes | 100.0% |
| 7 | Lens | 96.9% |
| 8 | Lacrimal glands | 69.8% |
| 9 | Optic nerves | 63.5% |
| 10 | Cochleae | 89.4% |
| 11 | Spinal cord | 59.6% |
| 12 | Chiasm | 63.3% |
| 13 | Brainstem | 43.6% |
| 14 | Temporal lobe | 82.0% |
| 15 | Submandibular | 92.5% |
| 16 | Lips | 26.7% |
| 17 | Esophagus | 55.2% |
| 18 | Pituitary gland | 100.0% |

| 19 | Thyroid gland | 96.3% |
|----|---------------|-------|
| 20 | Larynx        | 56.0% |
| 21 | Oral cavity   | 43.8% |
| 22 | TMJ           | 66.1% |

Next, we clustered 1175 contours into six structure groups (K=6) by the k-means method with 5 PCs. The Λ value of this clustering was 76.9%. Most of the 22 structures could not be classified into a single group, but the contours of one structure belonged to several subgroups. This can be seen in Figure 9, where the $P_{i,k}$ of anatomical structure, i and the k-th subgroup defined by Eq. (1) are plotted using bar graphs for all 22 structures. It is noteworthy that structure #19 had all contours in subgroup 1 (red). The contours of structure #5 solely belonged to subgroup 2 (orange). Most of the contours of structures #6, #13, and #14 belonged to subgroup 3 (coral). The contours of structures #7 and 18 were in subgroup 4 (amber). Meanwhile, GTV was distributed over all six subgroups.

Figure 9: Probability distributions over six subgroups of 22 anatomical structures

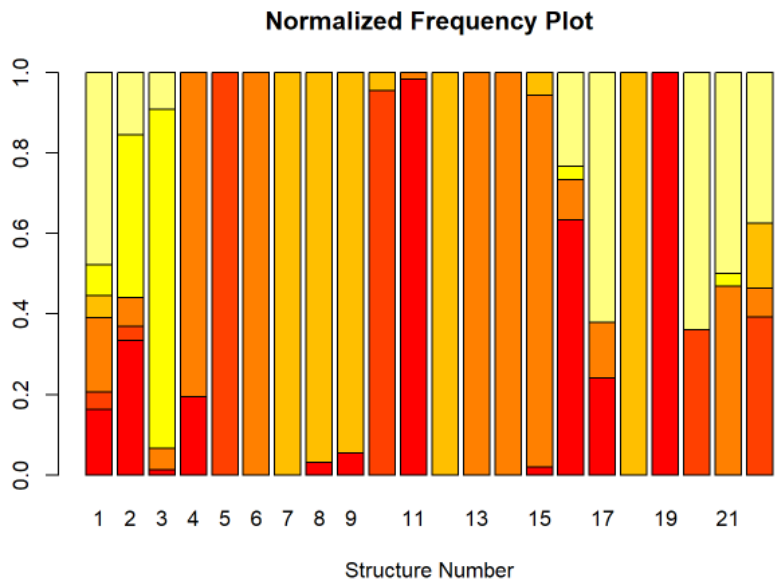

Stage-2

To classify 22 anatomical structures based on the data presented in Figure 9, we used the hierarchical clustering method (Stage 2 clustering). Here, 22 anatomical structures were assigned to one of six structure groups. The clustering result was expressed by a dendrogram shown in Figure 10. We can recognize six structure groups, considering the number of tree branches at the 3rd level indicated by a red horizontal line. Table 7 summarizes the clustering result where 22 anatomical structures were assigned to one of the six structure groups. The results show structures #1 (GTV), #17 (Esophagus), #20 (Larynx), #21 (Oral cavity), and #22 (TMJ) made up one structure group. Structures #5 (Mandible) and #10 (Cochleae), which contained bony tissue, could represent another structure group. Structures #2 (CTV) and #3 (PTV) formed the structure group 2. Furthermore, the dendrogram of Figure 10 tells that the parotids (#4) and submandibular (#15), which are salivary glands, were in the same structure group.

Table 7: Classification of 22 structure groups into six structure groups, which were obtained from level 3 of the dendrogram in Figure 10 (5 PCs)

| Structure group | Group members | Avg. CT# [HU] |
|---|---|---|
| 1 | 7, 8, 9, 12, 18 | $3.1 \pm 1.33$ |
| 2 | 5, 10 | $32.1 \pm 6.13$ |
| 3 | 4, 6, 13, 14, 15 | $4.2 \pm 1.47$ |
| 4 | 1, 17, 20, 21, 22 | $25.0 \pm 4.55$ |
| 5 | 11, 16, 19 | $10.9 \pm 6.55$ |
| 6 | 2, 3 | $33.6 \pm 8.09$ |

Figure 10: Cluster dendrogram

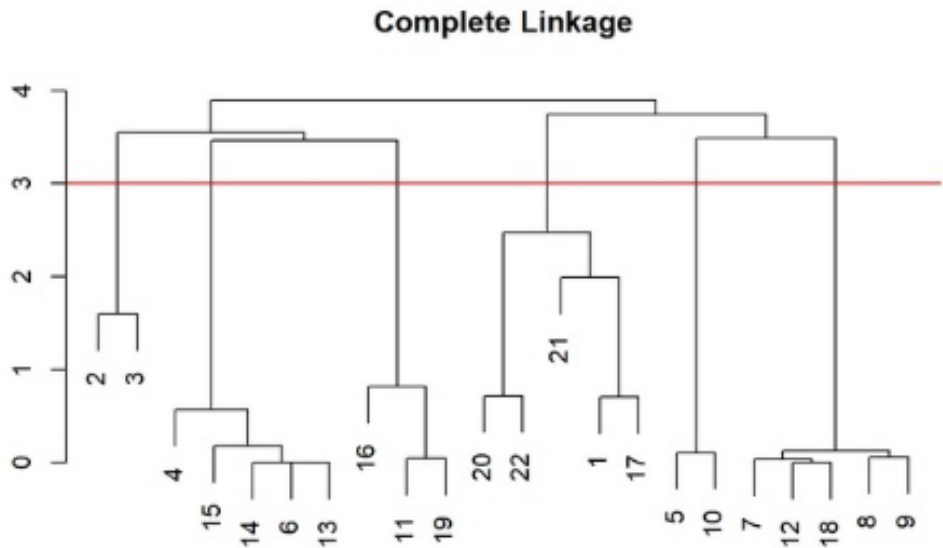

**Discussion**

The radiomics features of structures containing tumors, i.e., GTV, CTV, and PTV, were noticeably different from other structures containing only healthy tissue. In particular, CTV and PTV had common features distinguishable from others, as indicated by the formation of one structure group after the two-stage clustering.

The clustering using radiomics features successfully differentiated contours belonging to some anatomical structures. Among the 22 structures that we examined in this study, the contours of eight structures were grouped in the same subgroup with higher than 80% accuracy. Meanwhile, the algorithm failed with GTV, CTV, PTV, and Lips with an accuracy lower than 30%.

This study developed a two-stage clustering method to assign 1175 contours into structure groups. In Stage 1, the 1175 contours were clustered into K groups using five principal components derived from 174 radiomics features. The analysis showed that K=22 lead to sufficiently accurate clustering of contours into 22 structure groups. Furthermore, after clustering 1175 contours into six subgroups (K=6), we did the second clustering (Stage-2 clustering) to cluster 22 anatomical structures into six structure groups. The resulting cluster dendrogram showed that GTV, esophagus, and oral cavity belonged to the same structure group. CTV and PTV formed one independent structure group because these contours contain the GTV volume and surrounding tissue, including bony structures. Two bony anatomical structures of the mandible and cochleae formed one structure group. Three other structure groups consisted of several anatomic structures with some common features. The results could be partially explained by examining the anatomical structures' mass density. However, even the density is not only the characteristics common to the structures grouped in the same structure group. For example, as seen in Table 7, the average CT number of structure groups 2, 4, and 6 are similar, but those are created with different anatomical structures.

In a recent review paper, Larue et al. (33) briefly discusses applications of radiomics to normal tissue. The primary utility of normal tissue radiomics is to assess the change in normal tissue to injuries related to cancer treatment and development using a series of imaging studies. The current study opens a potentially new direction. One can use radiomics features strongly associated with the physiological and biological characteristics of normal tissue or organs to detect functional changes of the organs through images.

In this study, we used an unsupervised algorithm to classify anatomical contours. We can also use supervised ML algorithms such as support vector machines or even convolution neural networks to classify the contours into predefined categories and generate models for predicting the anatomical location from the CT images. Such a study could identify important radiomics features specific to each structure category.

**Conclusions**

We showed that the CT radiomics exhibited recognizable differences among twenty-two anatomical structures in head and neck areas. The novel two-stage clustering algorithm could categorize the anatomical structures into subgroups, sharing common biological functions and tissue types. In conclusion, our results implied that the radiomics of CT images could help discover the differences of biological processes, microenvironment, and cellular structures among anatomical structures in the head and neck.

Appendix-A: Complete list of radiomics features used for the current study

[Available upon request to the authors]

**Author's Note**

This study received ethical approval from the All India Institute of Medical Sciences, Institute Ethics Committee: IEC-87/07.02.2020, RP-04/2020. This is an IRB-approved retrospective study. All patient information was de-identified, and patient consent was not required.

**Declaration of Conflicting Interests**

The authors declared no potential conflicts of interest for this article's research, authorship, and publication.

**Funding**

The authors received no financial support for the research, authorship, and publication of this article.
25

**References**

1. Fried DV, Tucker SL, Zhou S, Liao Z, Mawlawi O, Ibbott G, et al. Prognostic value and reproducibility of pretreatment CT texture features in stage III non-small cell lung cancer. Int J Radiat Oncol Biol Phys. 2014;90(4):834-42.

2. Cunliffe A, Armato SG, 3rd, Castillo R, Pham N, Guerrero T, Al-Hallaq HA. Lung texture in serial thoracic computed tomography scans: correlation of radiomics-based features with radiation therapy dose and radiation pneumonitis development. Int J Radiat Oncol Biol Phys. 2015;91(5):1048-56.

3. Parmar C, Leijenaar RT, Grossmann P, Rios Velazquez E, Bussink J, Rietveld D, et al. Radiomic feature clusters and prognostic signatures specific for Lung and Head & Neck cancer. Sci Rep. 2015;5:11044.

4. Coroller TP, Agrawal V, Narayan V, Hou Y, Grossmann P, Lee SW, et al. Radiomic phenotype features predict pathological response in non-small cell lung cancer. Radiotherapy and oncology : journal of the European Society for Therapeutic Radiology and Oncology. 2016;119(3):480-6.

5. van Timmeren JE, Leijenaar RTH, van Elmpt W, Reymen B, Oberije C, Monshouwer R, et al. Survival prediction of non-small cell lung cancer patients using radiomics analyses of cone-beam CT images. Radiother Oncol. 2017;123(3):363-9.

6. Yu W, Tang C, Hobbs BP, Li X, Koay EJ, Wistuba, II, et al. Development and Validation of a Predictive Radiomics Model for Clinical Outcomes in Stage I Non-small Cell Lung Cancer. Int J Radiat Oncol Biol Phys. 2018;102(4):1090-7.

7. Shi L, He Y, Yuan Z, Benedict S, Valicenti R, Qiu J, et al. Radiomics for Response and Outcome Assessment for Non-Small Cell Lung Cancer. Technology in cancer research & treatment. 2018;17:1533033818782788-.

8. Spraker MB, Wootton LS, Hippe DS, Ball KC, Peeken JC, Macomber MW, et al. MRI Radiomic Features Are Independently Associated With Overall Survival in Soft Tissue Sarcoma. Adv Radiat Oncol. 2019;4(2):413-21.

9. Hirose TA, Arimura H, Ninomiya K, Yoshitake T, Fukunaga JI, Shioyama Y. Radiomic prediction of radiation pneumonitis on pretreatment planning computed tomography images prior to lung cancer stereotactic body radiation therapy. Sci Rep. 2020;10(1):20424.

10. Zhai T-T, Langendijk JA, van Dijk LV, van der Schaaf A, Sommers L, Vemer-van den Hoek JGM, et al. Pre-treatment radiomic features predict individual lymph node failure for head and neck cancer patients. Radiotherapy and Oncology. 2020;146:58-65.

11. Kawahara D, Tang X, Lee CK, Nagata Y, Watanabe Y. Predicting the Local Response of Metastatic Brain Tumor to Gamma Knife Radiosurgery by Radiomics With a Machine Learning Method. Frontiers in oncology. 2020;10(3003):569461.

12. Mouraviev A, Detsky J, Sahgal A, Ruschin M, Lee YK, Karam I, et al. Use of radiomics for the prediction of local control of brain metastases after stereotactic radiosurgery. Neuro Oncol. 2020;22(6):797-805.

13. Tanadini-Lang S, Balermpas P, Guckenberger M, Pavic M, Riesterer O, Vuong D, et al. Radiomic biomarkers for head and neck squamous cell carcinoma. Strahlentherapie und Onkologie. 2020;196(10):868-78.

14. Able H, Wolf-Ringwall A, Rendahl A, Ober CP, Seelig DM, Wilke CT, et al. Computed tomography radiomic features hold prognostic utility for canine lung tumors: An analytical study. PLOS ONE. 2021;16(8):e0256139.

15. Hsu CY, Doubrovin M, Hua CH, Mohammed O, Shulkin BL, Kaste S, et al. Radiomics Features Differentiate Between Normal and Tumoral High-Fdg Uptake. Sci Rep. 2018;8(1):3913.

16. Jiang X, Xie F, Liu L, Peng Y, Cai H, Li L. Discrimination of malignant and benign breast masses using automatic segmentation and features extracted from dynamic contrast-enhanced and diffusion-weighted MRI. Oncol Lett. 2018;16(2):1521-8.

17. Aerts HJ, Velazquez ER, Leijenaar RT, Parmar C, Grossmann P, Carvalho S, et al. Decoding tumour phenotype by noninvasive imaging using a quantitative radiomics approach. Nat Commun. 2014;5:4006.

18. Li H, Zhu Y, Burnside ES, Huang E, Drukker K, Hoadley KA, et al. Quantitative MRI radiomics in the prediction of molecular classifications of breast cancer subtypes in the TCGA/TCIA data set. NPJ Breast Cancer. 2016;2:16012.

19. Peeken JC, Bernhofer M, Spraker MB, Pfeiffer D, Devecka M, Thamer A, et al. CT-based radiomic features predict tumor grading and have prognostic value in patients with soft tissue sarcomas treated with neoadjuvant radiation therapy. Radiother Oncol. 2019;135:187-96.

20. Romeo V, Cuocolo R, Ricciardi C, Ugga L, Cocozza S, Verde F, et al. Prediction of Tumor Grade and Nodal Status in Oropharyngeal and Oral Cavity Squamous-cell Carcinoma Using a Radiomic Approach. Anticancer Res. 2020;40(1):271-80.

21. Kobayashi K, Miyake M, Takahashi M, Hamamoto R. Observing deep radiomics for the classification of glioma grades. Sci Rep. 2021;11(1):10942.

22. Lee SH, Cho HH, Kwon J, Lee HY, Park H. Are radiomics features universally applicable to different organs? Cancer imaging : the official publication of the International Cancer Imaging Society. 2021;21(1):31.

23. Rosenstein BS, West CM, Bentzen SM, Alsner J, Andreassen CN, Azria D, et al. Radiogenomics: radiobiology enters the era of big data and team science. Int J Radiat Oncol Biol Phys. 2014;89(4):709-13.

24. Socarras Fernandez JA, Monnich D, Leibfarth S, Welz S, Zwanenburg A, Leger S, et al. Comparison of patient stratification by computed tomography radiomics and hypoxia positron emission tomography in head-and-neck cancer radiotherapy. Phys Imaging Radiat Oncol. 2020;15:52-9.

25. Tunali I, Tan Y, Gray JE, Katsoulakis E, Eschrich SA, Saller J, et al. Hypoxia-related radiomics predict immunotherapy response: A multi-cohort study of NSCLC. bioRxiv. 2020:2020.04.02.020859.

26. Wang JH, Wahid KA, van Dijk LV, Farahani K, Thompson RF, Fuller CD. Radiomic biomarkers of tumor immune biology and immunotherapy response. Clinical and Translational Radiation Oncology. 2021;28:97-115.

27. Bettinelli A, Branchini M, De Monte F, Scaggion A, Paiusco M. Technical Note: An IBEX adaption toward image biomarker standardization. Medical physics. 2020;47(3):1167-73.

28. Zwanenburg A, Leger S, Vallières M, Löck S. Image biomarker standardisation initiative (IBSI). arXiv [Internet]. 2019 December 01, 2016:[arXiv:1612.07003 p.]. Available from: https://ui.adsabs.harvard.edu/abs/2016arXiv161207003Z.

29. James G, Witten D, Hastie T, Tibshirani R. An Introduction to Statistical Learning with Applications in R. New York, NY: Springer; 2017.

30. Monti S, Tamayo P, Mesirov J, Golub T. Consensus Clustering: A Resampling-Based Method for Class Discovery and Visualization of Gene Expression Microarray Data. Machine Learning. 2003;52(1):91-118.

31. Wilkerson MD, Hayes DN. ConsensusClusterPlus: a class discovery tool with confidence assessments and item tracking. Bioinformatics. 2010;26(12):1572-3.

32. Team RC. R: A language and environment for statistical computing. Vienna, Austria: R Foundation for Statistical Computing; 2003.

33. Larue RT, Defraene G, De Ruysscher D, Lambin P, van Elmpt W. Quantitative radiomics studies for tissue characterization: a review of technology and methodological procedures. The British journal of radiology. 2017;90(1070):20160665.

**Figure captions**

Figure 1: CT images of a typical patient with the contours of GTV, right and left parotids, and mandible.

Figure 2: Process map of the analysis procedure

Figure 3: Heatmap. 1357 anatomical structure in the horizontal axis and 174 radiomic features in the vertical axis.

Figure 4: Correlation of radiomic feature values

Figure 5: Total within clusters sum of squares (TWCCS) vs. the number of clusters

Figure 6: 1175 contours plotted for the axes with the first two principal components

Figure 7: Scree plot: Percent Variation vs. the number of principal components

Figure 8: Accuracy of clustering of 1175 contours of 22 anatomical structures to 22 subgroups

Figure 9: Probability distributions over six subgroups of 22 anatomical structures

Figure 10: Cluster dendrogram